\documentclass[doublecol]{epl2}
\usepackage{amssymb,amsmath,graphicx,setspace,color,rotating,url}
\title{The components of empirical multifractality in financial returns}

\author{Wei-Xing Zhou\inst{1,2,3,4}\footnote{e-mail: wxzhou@ecust.edu.cn}}
\shortauthor{W.-X. Zhou}

\institute{
  \inst{1} School of Business, East China University of Science and Technology, Shanghai 200237, China\\
  \inst{2} School of Science, East China University of Science and Technology, Shanghai 200237, China\\
  \inst{3} Research Center for Econophysics, East China University of Science and Technology, Shanghai 200237, China\\
  \inst{4} Research Center on Fictitious Economics \& Data Science, Chinese Academy of Sciences, Beijing 100080, China%
 }


\abstract{We perform a systematic investigation on the components of the  empirical multifractality of financial returns using the daily data of Dow Jones Industrial Average from 26 May 1896 to 27 April 2007 as an example. The temporal structure and fat-tailed distribution of the returns are considered as possible influence factors. The multifractal spectrum of the original return series is compared with those of four kinds of surrogate data: (1) shuffled data that contain no temporal correlation but have the same distribution, (2) surrogate data in which any nonlinear correlation is removed but the distribution and linear correlation are preserved, (3) surrogate data in which large positive and negative returns are replaced with small values, and (4) surrogate data generated from alternative fat-tailed distributions with the temporal correlation preserved. We find that all these factors have influence on the multifractal spectrum. We also find that the temporal structure (linear or nonlinear) has minor impact on the singularity width $\Delta\alpha$ of the multifractal spectrum while the fat tails have major impact on $\Delta\alpha$, which confirms the earlier results. In addition, the linear correlation is found to have only a horizontal translation effect on the multifractal spectrum in which the distance is approximately equal to the difference between its DFA scaling exponent and 0.5. Our method can also be applied to other financial or physical variables and other multifractal formalisms.}

\begin{document}

\maketitle

\section{Introduction}

There are a wealth of studies showing that financial markets exhibit
multifractal nature
\cite{Ghashghaie-Breymann-Peinke-Talkner-Dodge-1996-Nature,Mantegna-Stanley-1996-Nature,Mandelbrot-1999-SA}.
Many different methods have been applied to characterize the hidden
multifractal behavior in finance, for instance, the fluctuation
scaling analysis
\cite{Eisler-Kertesz-Yook-Barabasi-2005-EPL,Eisler-Kertesz-2007-EPL,Jiang-Guo-Zhou-2007-EPJB},
the structure function (or height-height correlation function)
method
\cite{Ghashghaie-Breymann-Peinke-Talkner-Dodge-1996-Nature,Vandewalle-Ausloos-1998-EPJB,Ivanova-Ausloos-1999-EPJB,Schmitt-Schertzer-Lovejoy-1999-ASMDA,Schmitt-Schertzer-Lovejoy-2000-IJTAF,Calvet-Fisher-2002-RES,Ausloos-Ivanova-2002-CPC,Gorski-Drozdz-Speth-2002-PA,AlvarezRamirez-Cisneros-IbarraValdez-Soriano-2002-PA,Balcilar-2003-EMFT,Lee-Lee-2005a-JKPS,Lee-Lee-Rikvold-2006-PA},
the multiplier method \cite{Jiang-Zhou-2007-PA}, the multifractal
detrended fluctuation analysis (MF-DFA)
\cite{Kantelhardt-Zschiegner-KoscielnyBunde-Havlin-Bunde-Stanley-2002-PA,Matia-Ashkenazy-Stanley-2003-EPL,Kwapien-Oswiecimka-Drozdz-2005-PA,Lee-Lee-2005b-JKPS,Oswiecimka-Kwapien-Drozdz-2005-PA,Jiang-Ma-Cai-2007-PA,Lee-Lee-2007-PA,Lim-Kim-Lee-Kim-Lee-2007-PA,Su-Wang-Huang-2009-JKPS},
the partition function method
\cite{Sun-Chen-Wu-Yuan-2001-PA,Sun-Chen-Yuan-Wu-2001-PA,Ho-Lee-Wang-Chuang-2004-PA,Wei-Huang-2005-PA,Gu-Chen-Zhou-2007-EPJB,Du-Ning-2008-PA,Zhuang-Yuan-2008-PA,Wei-Wang-2008-PA,Zhou-2010-cnJMSC,Jiang-Zhou-2008a-PA,Su-Wang-2009-JKPS},
the wavelet transform approaches
\cite{Struzik-Siebes-2002-PA,Turiel-Perez-Vicente-2003-PA,Turiel-Perez-Vicente-2005-PA,Oswiecimka-Kwapien-Drozdz-Rak-2005-APPB},
to list a few. There are also efforts seeking for applications of
the extracted multifractal spectra. Some researchers reported that
the observed multifractal singularity spectrum has predictive power
for price fluctuations
\cite{Sun-Chen-Yuan-Wu-2001-PA,Wei-Huang-2005-PA,Su-Wang-2009-JKPS},
can serve as a measure of market risk by introducing a new concept
termed {\em{multifractal volatility}} \cite{Wei-Wang-2008-PA},
and can be used to quantify the inefficiency of markets
\cite{Zunino-Tabak-Figliola-Perez-Garavaglia-Rosso-2008-PA}.

An important and subtle issue of multifractality is about its
origin. An even critical question is to ask whether the extracted
multifractality is intrinsic or apparent. Indeed, it has been shown
that an exact monofractal financial model can lead to an artificial
multifractal behavior \cite{Bouchaud-Potters-Meyer-2000-EPJB}. It is
usually argued in the Econophysics community that the sources of
multifractal nature in financial time series are the fat tails
and/or the long-range temporal correlation
\cite{Kantelhardt-Zschiegner-KoscielnyBunde-Havlin-Bunde-Stanley-2002-PA}.
However, possessing long memory is not sufficient for the presence
of multifractality and one has to have a nonlinear process with
long-memory in order to have multifractality
\cite{Saichev-Sornette-2006-PRE}. In many cases, the null hypothesis
that the reported multifractal nature is stemmed from the large
price fluctuations cannot be rejected \cite{Lux-2004-IJMPC}.

In this Letter, we focus on the multifractal detrended fluctuation
analysis of financial logarithmic returns defined as
\begin{equation}
 r(t) = \ln[P(t)/P(t-1)]
 \label{Eq:rt}
\end{equation}
where $P(t)$ is the price at time $t$. Specifically, we use the
daily data of the Dow Jones Industrial Average (DJIA) from 26 May
1896 to 27 April 2007 (totally 30147 trading days) to illustrate the
method and results. The reason is simply that most studies in this
direction use MF-DFA on stock returns. Nevertheless, the methodology
is quite general and also applies in the study of other financial
variables and other multifractal analysis.

The most studied factor is the temporal correlation in the return
series, where the singularity spectrum of the real data is compared
with that of the randomly shuffled data
\cite{Matia-Ashkenazy-Stanley-2003-EPL,Oswiecimka-Kwapien-Drozdz-Rak-2005-APPB,Lee-Lee-2005b-JKPS,Kwapien-Oswiecimka-Drozdz-2005-PA,Kumar-Deo-2009-PA,deSouza-Queiros-2009-CSF}.
All these studies show that the shuffled data have non-shrinking
singularity width $\Delta\alpha$ and the vertex $(\alpha,f(\alpha))$
with $q=0$ may shift left more or less. These observations imply
that the heavy-tailed distribution of the returns has crucial impact
on the singularity width. Extensive numerical experiments using
uncorrelated time series obeying $q$-Gaussian distributions with
different tail exponents unveil a convergence to monofractalilty in
the Gaussian attraction basin and to bifractality in the L\'evy
attraction basin \cite{Drozdz-Kwapien-Oswiecimka-Speth-2009-XXX},
which is consistent with the analytic derivation of truncated L\'evy
flights \cite{Nakao-2000-PLA}. Similar phenomena are observed for
exponential distributions in the partition function framework
\cite{vonHardenberg-Thieberger-Provenzale-2000-PLA}. To understand
the impact of the distribution, one can either remove the large
positive and negative returns
\cite{Oh-Eom-Havlin-Jung-Wang-Stanley-Kim-2010-PRE} or generate
surrogate data having a Gaussian distribution while keeping the
linear correlation of the original data
\cite{Norouzzadeh-Rahmani-2006-PA,Lim-Kim-Lee-Kim-Lee-2007-PA,Su-Wang-Huang-2009-JKPS}.
In this Letter, we will systematically investigate these factors
together with a new factor reflecting possible hidden patterns in
the raw time series.

\section{Memory effect}

We adopt the MF-DFA method to obtain empirically the singularities
$\alpha$ and the corresponding spectrum $f(\alpha)$ for each time
series
\cite{Kantelhardt-Zschiegner-KoscielnyBunde-Havlin-Bunde-Stanley-2002-PA}. In
all the cases, the scaling range of the detrended fluctuation
function with respect to the time scale is $[30, 3000]$, the moment
order $q$ varies from -5 to 5, and the second-order polynomial
(parabolic) is used for detrending.The determination of
the scaling range is a subtle issue since the intrinsic scaling
behavior will be masked or deformed by high-order trends or
nonlinearity
\cite{Hu-Ivanov-Chen-Carpena-Stanley-2001-PRE,Chen-Ivanov-Hu-Stanley-2002-PRE,Chen-Hu-Carpena-Bernaola-Galvan-Stanley-Ivanov-2005-PRE}.
The power laws of the detrended fluctuation functions are not
perfect. However, we find that the resulting multifractal spectra do
not change much when we using a much narrower scaling range
$[30, 300]$. When the time scale is 3000, there are only 20 boxes
from both ends of the time series. To gain better statistics, we
modify slightly the MF-DFA algorithm in the partitioning of boxes for a
given time scale. Consider a time series $\{r(t)| t=1,2,\cdots,N\}$.
For time scale $s$, we select a random sequence
$\{j_i| i=1,2,\cdots,n\}$, which are uniformly distributed in
$[1,t-s+1]$. In our analysis, we use $n=2000$. This sequence
determines $n$ boxes $[j_i,j_i+s-1]$, in which the locally detrended
fluctuation functions $F_i(s)$ are calculated. If the time series is
long-term power-law correlated, $F(s)$ scales as a power law of $s$
\cite{Kantelhardt-Zschiegner-KoscielnyBunde-Havlin-Bunde-Stanley-2002-PA}
\begin{equation}
 F(s) = \left\{\frac{1}{n}\sum_{i=1}^n [F_i(s)]^q\right\}^{1/q}
      \sim s^{h(q)}
 \label{Eq:Fs:hq}
\end{equation}
The mass scaling exponent $\tau(q)$ in the partition function
formulism can be determined as
\begin{equation}
 \tau(q) = qh(q)-1
 \label{Eq:tau:q}
\end{equation}
and the singularity strength $\alpha$ and its spectrum $f(\alpha)$
can be calculated according to the Legendre transform
\cite{Halsey-Jensen-Kadanoff-Procaccia-Shraiman-1986-PRA}
\begin{equation}
\left\{
 \begin{array}{lcc}
 \alpha = h(q)-q h'(q) \\
 f(\alpha) = q[\alpha-h(q)]+1
 \end{array}
 \label{Eq:alpha:f}
\right..
\end{equation}

\begin{figure}[htb]
\centering
\includegraphics[width=7cm]{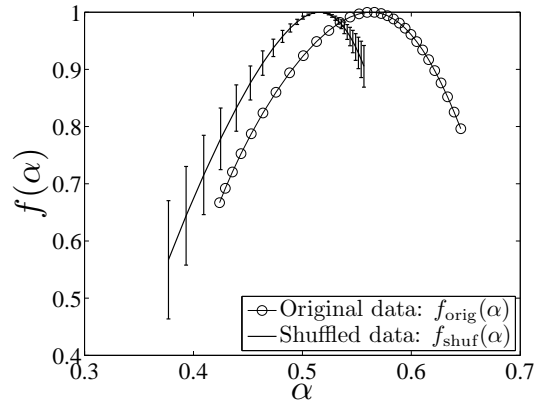}
\caption{\label{Fig:MFComp:Memory} Comparison of the multifractal
spectra $f_{\rm{orig}}(\alpha)$ for the original data and
$f_{\rm{shuf}}(\alpha)$ for the shuffled data.}
\end{figure}

The multifractal spectrum $f_{\rm{orig}}(\alpha)$ of the return
series is thus determined and shown in fig.~\ref{Fig:MFComp:Memory}.
We then shuffle the return series 100 times and determine their
singularity spectra $f_{\rm{shuf}}(\alpha)$. For each point on the
$f_{\rm{shuf}}(\alpha)$ curve of the shuffled data, $\alpha$ and
$f_{\rm{shuf}}(\alpha)$ are the arithmetic averages of the
respective 100 values of the shuffled data and the error bar is the
corresponding standard deviation. The singularity width is
$\Delta\alpha_{\rm{orig}}=0.22$ for the original data and
$\Delta\alpha_{\rm{shuf}}=0.18\pm0.04$ for the shuffled data. These
results are consistent with previous works. We also observe that the
surrogate data have negative dimensions for larger $q$, which implies that the
surrogate data contains more randomness in large values of local
detrended fluctuations \cite{Mandelbrot-1990a-PA,Mandelbrot-1991-PRSA}.

\section{Effect of distribution}

As we have shown above that, shuffling the return time series does
not eliminate its multifractality. This is not surprising since
there is no (or very weak) long-term correlation in the returns.
Since the shuffled time series still exhibits a wide singularity
spectrum, it is natural to conjecture that the distribution of
returns has essential impact on the $f(\alpha)$ curve. We try to
systematically test this conjecture using surrogate data. Two
methods are adopted to generate surrogate data. One method is to
replace the returns having large magnitudes with random numbers
drawn from a normal distribution
\cite{Oh-Eom-Havlin-Jung-Wang-Stanley-Kim-2010-PRE}. The other
method is to substitute the raw returns with data drawn from
prescribed distributions by keeping the ranking order which is
relevant to the phase randomization algorithm
\cite{Theiler-Eubank-Longtin-Galdrikian-Farmer-1992-PD} but with
some differences.

\subsection{The truncation method}

The truncation method was originally proposed to study the impact of
large positive and negative values on the multifractal singularity
width $\Delta\alpha$ of the foreign exchange rate returns, where the
returns with the magnitudes greater than $M\sigma$ were eliminated
and replaced by linear interpolations
\cite{Oh-Eom-Havlin-Jung-Wang-Stanley-Kim-2010-PRE}, where $\sigma$
is the standard deviation of the raw time series. For convenience,
the resulting data are termed as truncated data. For the FX returns,
the singularity width $\Delta\alpha$ of the truncated data increases
as the normalized threshold $M$ increases
\cite{Oh-Eom-Havlin-Jung-Wang-Stanley-Kim-2010-PRE}. When the
truncated time series is shuffled, the singularity width
$\Delta\alpha$ decreases dramatically
\cite{Oh-Eom-Havlin-Jung-Wang-Stanley-Kim-2010-PRE}. These analyses
illustrate that large values in the FX returns have significant
impact on the width of multifractal singularity and the temporal
structure of the truncated series becomes a stronger factor on its
multifractality.

\begin{figure}[htb]
 \centering
 \includegraphics[width=7cm]{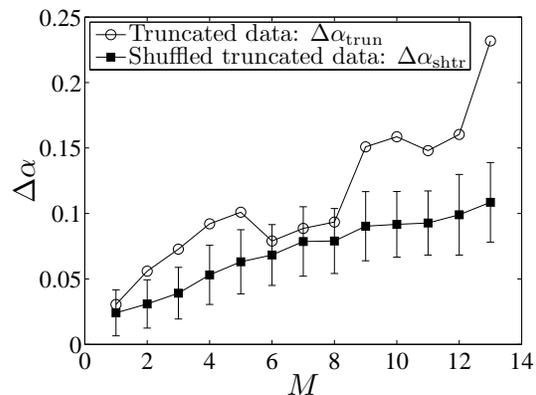}
 \caption{Dependence of the singularity width $\Delta\alpha$
 on the normalized threshold $M$ for the truncated time series from which returns with $|r(t)|>M\sigma$ have been
deleted and replaced ($\Delta\alpha_{\rm{trun}}$) and for the
shuffled truncated data ($\Delta\alpha_{\rm{shtr}}$). }
 \label{Fig:MFComp:PDF:Trancate:Dalpha}
\end{figure}

We have followed this idea and constructed the truncated time series
in a slightly different way. The substitute for returns with
$|r(t)|>M\sigma$ is a collection of returns re-sampled randomly from
the return series with $|r(t)|\leqslant M\sigma$. We have generated
100 truncated data sets with the threshold $M\sigma$ spanning from
$\sigma$ to $13\sigma$. The dependence of the singularity width
$\Delta\alpha_{\rm{trun}}$ on $M$ is illustrated in
fig.~\ref{Fig:MFComp:PDF:Trancate:Dalpha}. It is clear that the
width of the multifractal spectrum shrinks when the normalized
threshold $M$ decreases. For each vale of $M$, the truncated data
are shuffled to generate 100 shuffled truncated data sets. For each
shuffled truncated data set, we determine its singularity width
$\Delta\alpha_{\rm{shtr}}$. The results are also presented in
fig.~\ref{Fig:MFComp:PDF:Trancate:Dalpha} for comparison. On
average, we find that
$\Delta\alpha_{\rm{shtr}}<\Delta\alpha_{\rm{trun}}$ with some
exceptions where $\Delta\alpha_{\rm{shtr}}=\Delta\alpha_{\rm{trun}}$
within the error bars. This phenomenon is similar to the case of FX
returns \cite{Oh-Eom-Havlin-Jung-Wang-Stanley-Kim-2010-PRE}.

\subsection{Surrogate data with different PDFs}

An alternative method to generate surrogate data is to replace the
raw data by random numbers drawn from a prescribed distribution. A
similar idea was implemented to investigate the sources of
multifractality in the returns of the Iranian rial - US dollar
exchange rate \cite{Norouzzadeh-Rahmani-2006-PA}, where the phase
randomization algorithm was adopted and the surrogate data have a
Gaussian distribution while keeping the linear correlation of the
original data
\cite{Theiler-Eubank-Longtin-Galdrikian-Farmer-1992-PD}. Our
algorithm for generating surrogate data is described as follows. For
a given distribution, we generate a sequence of random numbers
$\{x_0(t)| t=1,2,\cdots,T\}$, which are rearranged such that the
rearranged series $\{x(t)| t=1,2,\cdots,T\}$ has the same rank
ordering as the return series $\{r(t)| t=1,2,\cdots,T\}$. In other
words, $x(t)$ should rank $n$ in sequence $\{x(t)| t=1,2,\cdots,T\}$
if and only if $r(t)$ ranks $n$ in the $\{r(t)| t=1,2,\cdots,T\}$
sequence \cite{Bogachev-Eichner-Bunde-2007-PRL,Zhou-2008-PRE}. The
series $\{x(t)\}$ is rescaled to have the same standard deviation
$\sigma$ of the returns $\{r(t)\}$:
\begin{equation}
 x(t) \to x(t)\times\sigma/\sigma_x+\mu,
 \label{Eq:xt:rescale}
\end{equation}
where $\sigma_x$ is the standard deviation of $\{x(t)\}$ and $\mu$
is the sample mean of $\{r(t)\}$.

In our analysis, we have investigated two types of distributions.
The first one is a family of ``double'' Weibull distributions
\begin{subequations}
\begin{equation}
 p(x) = \beta x^{\beta-1}e^{-|x-\mu|^\beta},
 \label{Eq:pdf:Weibull}
\end{equation}
where the shape parameter $\beta$ describes the heaviness of the
tails and we require that $\beta<1$. The second one is a family of
Student's t distributions
\begin{equation}
 p(x) = \frac{\Gamma\left(\frac{\gamma+1}{2}\right)}{\sqrt{\gamma\pi}\Gamma(\frac{\gamma}{2})}
        \left[1+\frac{(x-\mu)^2}{\gamma}\right]^{-(\gamma+1)/2},
 \label{Eq:pdf:Student}
\end{equation}
which have power-law tails with exponent $\gamma$.
\end{subequations}

\begin{figure}[h!]
 \centering
 \includegraphics[width=6cm]{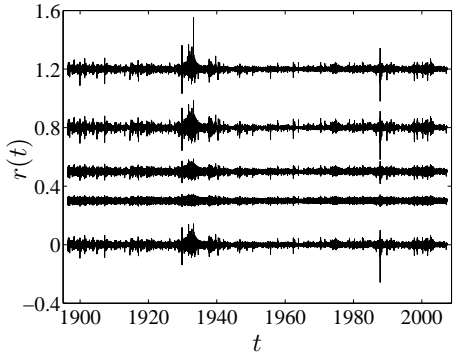}
 \caption{\label{Fig:MFComp:PDF:Surrogate:Series} Comparison of the original return time series with its surrogates.
 The surrogate time series from bottom to top are generated from a normal distribution,
 a Laplace distribution, a Weibull distribution with $\beta=0.5$, and a Student distribution with $\gamma=3$.
 The original return series is at the bottom and the four surrogate
 series have been vertically shifted for better visibility.}
\end{figure}

Fig.~\ref{Fig:MFComp:PDF:Surrogate:Series} compares the original
return series with four surrogate time series generated from a
normal distribution, a Laplace distribution, a Weibull distribution
with $\beta=0.5$, and a Student distribution with $\gamma=3$,
respectively. We can see that all the surrogate series exhibit
similar clustering phenomenon as the original returns. In other
words, the volatility (absolute value) of the surrogate data has
long-term correlation, which is still true for all the cases. This
intriguing feature is very important since it is absent in the
surrogate data according to the phase randomization algorithm. In
addition, the two surrogate series with fat tails share more
similarity with the original returns than those from the normal and
Laplace distributions. Certainly, a close scrutiny of the time
series will unveil differences in their finer structure.

\begin{figure}[htb]
 \centering
 \includegraphics[width=7cm]{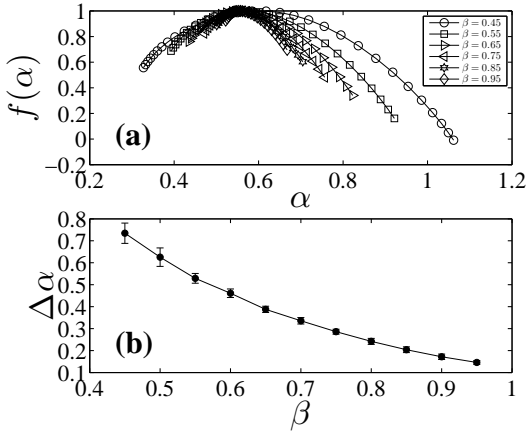}
 \caption{\label{Fig:MFComp:PDF:Surrogate:WBL} (a) Multifractal spectra of surrogate time
 series generated from Weibull distributions with $\beta=0.45$, 0.55, 0.65, 0.75, 0.85 and
 0.95, which have the same rank ordering as the original return series. (b) Dependence of the singularity width $\Delta\alpha$
 as a function of the exponent $\beta$. The error bars are the standard deviations for the 100 surrogate series.}
\end{figure}

For the case of Weibull distributions, we investigate 11 values of
the exponent $\beta$, varying from 0.45 to 0.95 with a spacing step
of 0.05. For smaller $\beta$ values, we find that the multifractal
spectra are not stable due to the poor statistics caused by extreme
jumps in the time series. For each $\beta$, we generate 100
surrogate time series and the average multifractal spectrum is
determined. The multifractal spectra for $\beta=0.45$, 0.55, 0.65,
0.75, 0.85 and 0.95 are illustrated in
fig.~\ref{Fig:MFComp:PDF:Surrogate:WBL}(a). It is evident that time
series with heavier tails (or small $\beta$) exhibits stronger
multifractality. This quantitative dependence of the average
singularity width $\Delta\alpha$ as a function of the exponent
$\beta$ is depicted in fig.~\ref{Fig:MFComp:PDF:Surrogate:WBL}(b).

\begin{figure}[htb]
 \centering
 \includegraphics[width=7cm]{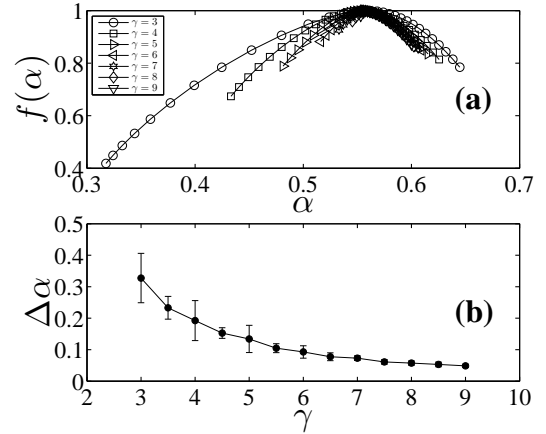}
 \caption{\label{Fig:MFComp:PDF:Surrogate:t} (a) Multifractal spectra of surrogate time
 series generated from Student's t distributions with $\gamma=3$, 4, 5, 6, 7, 8 and
 9, which have the same rank ordering as the original return series.
 Dependence of the singularity width $\Delta\alpha$
 as a function of the exponent $\beta$. The error bars are the standard deviations for the 100 surrogate series.}
\end{figure}

For the case of Student distributions, we investigate 13 values for
the tail exponent $\gamma$, varying from 3 to 9 with a spacing step
of 0.5. We choose the minimal value of $\gamma=3$ since the returns
at the transaction level are well modeled by Student's t
distribution with $\gamma=3$ \cite{Gu-Chen-Zhou-2008a-PA}, which is
the well-known inverse cubic law
\cite{Gopikrishnan-Meyer-Amaral-Stanley-1998-EPJB}, and the tail
exponent increases with the time scale
\cite{Ghashghaie-Breymann-Peinke-Talkner-Dodge-1996-Nature}. For
each $\gamma$, we generate 100 surrogate time series and the average
multifractal spectrum is determined. Figure
\ref{Fig:MFComp:PDF:Surrogate:t}(a) illustrates the multifractal
spectra for $\gamma=3$, 4, 5, 6, 7, 8 and 9. Again, it is evident
that time series with heavier tails (or small $\gamma$) exhibits
stronger multifractality. Comparing with
fig.~\ref{Fig:MFComp:PDF:Surrogate:WBL}(a), we find that the right
parts of the $f(\alpha)$ curves for the Student's t distributions
are closer to each other, which means that the small values
characterized by negative $q$'s are more irregular. The quantitative
dependence of the average singularity width $\Delta\alpha$ as a
function of the exponent $\gamma$ is shown in
fig.~\ref{Fig:MFComp:PDF:Surrogate:t}(b).

\section{Hidden nonlinear structure}

In the above analysis, we have investigated the impact of temporal
correlation and probability distribution on the multifractal nature.
More rigorously speaking, the shuffling approach is very aggressive,
which not only removes linear correlations but also eliminates any
hidden structure in the original return series. It is thus
interesting to assess the impact of hidden structure. This can be
done with the help of surrogate time series which has the same
distribution and linear temporal correlation as the original data.
We find that the DFA scaling exponent of the DJIA returns is $H=0.54$.
Although the DFA scaling exponent is very close to 0.5 for uncorrelated time
series, it may still contain some nontrivial information about the
linear correlation in the original series. It is worth noting that
this method will be more interesting for other financial time series
whose DFA scaling exponent is significantly larger than 0.5 and thus have
long memory.

The algorithm for the generating of surrogate data is based on a
simple iteration scheme called iterated amplitude-adjusted Fourier
transform (IAAFT) \cite{Schreiber-Schmitz-1996-PRL}, which is an
improved version of the phase randomization algorithm
\cite{Theiler-Eubank-Longtin-Galdrikian-Farmer-1992-PD}. The return
data $\{r(t)| t=1,2,\cdots,N\}$ are sorted resulting in a new
sequence $\{s_N\}$, and we obtain the squared amplitudes of the
Fourier transform of $\{s_N\}$, denoted as $\{S_k^2\}$. The initial
sequence $\{s_N^{(0)}\}$ of the iteration is a random shuffle of
$\{s_N\}$. In the $i$-th iteration, the squared amplitudes
$\{S_k^{2,(i)}\}$ of the Fourier transform of $\{s_N^{(0)}\}$ are
obtained and replaced by $\{S_k^2\}$, which are transformed back,
and then the resulting series are replaced by $\{s_N\}$ but keeping
the rank order. We generate 100 surrogate time series and perform
the iteration 20 times for each surrogate series.

\begin{figure}[htb]
\centering
\includegraphics[width=6.5cm]{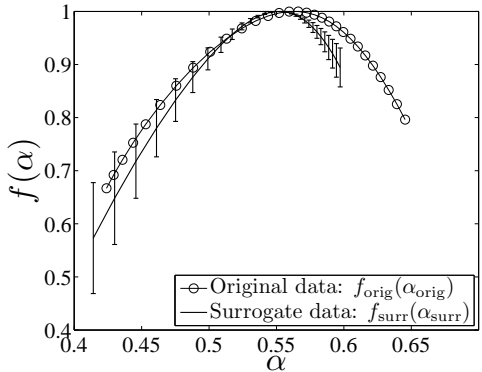}
\caption{\label{Fig:MFComp:Hidden} Comparison of the multifractal
spectra $f_{\rm{orig}}(\alpha)$ of the original data and
$f_{\rm{surr}}(\alpha)$ of the surrogate data.}
\end{figure}

Fig.~\ref{Fig:MFComp:Hidden} plots the averaged multifractal
spectrum $f_{\rm{surr}}(\alpha)$ of the surrogate series and the
$f_{\rm{orig}}(\alpha)$ curve of the original return series as well.
The singularity width is $\Delta\alpha_{\rm{surr}}=0.18\pm0.04$ for
the surrogate data, which is close to
$\Delta\alpha_{\rm{orig}}=0.22$ for the original data. For the right
part where $q>0$, the $f_{\rm{surr}}(\alpha)$ curve for surrogate
data is embraced by that for the original data. For the left part
where $q<0$, the two curves almost overlap with the error bar for a
large part. We also observe that the surrogate data have negative
dimensions, which implies that the surrogate data contain more
randomness in large values of local detrended fluctuations
\cite{Mandelbrot-1990a-PA,Mandelbrot-1991-PRSA}.

It is also interesting to compare the $f(\alpha)$ curve for the
surrogate data in fig.~\ref{Fig:MFComp:Hidden} and for the shuffled
data in fig.~\ref{Fig:MFComp:Memory}. We find that these two curves
$f_{\rm{surr}}(\alpha)$ and $f_{\rm{shuf}}(\alpha)$ have the same
shape with a horizontal shift. We have approximately that
\begin{equation}
 f_{\rm{surr}}(\alpha) =
 f_{\rm{shuf}}(\alpha + \Delta_\alpha),
 \label{Eq:f:surr:shuf}
\end{equation}
where $\Delta_\alpha = \alpha_{{\rm{surr}},q=0}- \alpha_{{\rm{shuf}},q=0} \approx 0.04$.
Note that the value of $\Delta_\alpha$ is close to the difference of
the two DFA scaling exponents, $H_{{\rm{surr}}}- H_{{\rm{shuf}}}=0.04$,
where $H_{{\rm{surr}}}=H_{\rm{orig}}=0.54$ and
$H_{{\rm{shuf}}}=0.50$. Since $\alpha(0)=h(0)$, this relation can be
understood as a first-order approximation in the sense that
$\alpha_{{\rm{surr}}}(0)- \alpha_{{\rm{shuf}}}(0) =
h_{{\rm{surr}}}(0)-h_{{\rm{shuf}}}(0) \approx
h_{{\rm{surr}}}(2)-h_{{\rm{shuf}}}(2) = H_{{\rm{surr}}}-
H_{{\rm{shuf}}}$. The relation (\ref{Eq:f:surr:shuf}) means that the
difference between the multifractal spectra of the surrogate and
shuffled data is caused by the weak temporal correlation remained in
the surrogate data. In general, the linear correlation in the time
series does not influence the shape of the singularity spectrum (and
thus its width) but rather has a shift effect.

To further quantify the presence of hidden nonlinear
features in the return time series and their effect on the
multifractal spectrum, some remarks are in order based on the
magnitude and sign decomposition method
\cite{Ashkenazy-Ivanov-Havlin-Peng-Goldberger-Stanley-2001-PRL,Ashkenazy-Havlin-Ivanov-Peng-SchulteFrohlinde-Stanley-2003-PA}.
The daily DJIA return series can be decomposed into a magnitude (or
volatility) series and a sign series. We find that the DFA scaling exponent
of the magnitude series is 0.87, confirming the well-known stylized
fact that stock market volatility has strong long memory. In
contrast, the magnitude series of the shuffled and the IAAFT surrogates are found to be
uncorrelated. These facts confirm the presence of nonlinearity in
the return time series. It is
well established that the phase-randomized surrogates of
heartbeat series have a vanishing width of singularity
\cite{Ivanov-Amaral-Goldberger-Havlin-Rosenblum-Struzik-Stanley-1999-Nature,Ivanov-Amaral-Goldberger-Havlin-Rosenblum-Stanley-Struzik-2001-Chaos}, which is due to the fact that the heartbeat time series do not have fat tails, but decay exponentially \cite{Ivanov-Rosenblum-Peng-Mietus-Havlin-Stanley-Goldberger-1998-PA}.

\section{Conclusions}

In summary, we have systematically studied the components of
empirical multifractality in financial returns based on shuffled and
surrogate data, taking the daily data of DJIA for more than one
hundred years as an example. We found that the distribution, the
linear correlation,  and the nonlinear structure have influence on
the singularity spectrum. When the degree of multifractality is
characterized by the singularity width $\Delta\alpha$, we found that
the temporal structure (linear correlation and nonlinearity) has
minor impact while the fat-tailed distribution plays a major role,
which is a confirmation of the earlier results.

\acknowledgments

I am grateful to Gao-Feng Gu, Zhi-Qiang Jiang and Guo-Hua Mu for
fruitful discussions. This work was partly supported by Shanghai
Educational Development Foundation (2008SG29) and the Program for
New Century Excellent Talents in University (NCET-07-0288).

\bibliography{E:/Papers/Auxiliary/Bibliography}

\begin{thebibliography}{10}
\expandafter\ifx\csname url\endcsname\relax\def\url#1{\texttt{#1}}\fi

\bibitem{Ghashghaie-Breymann-Peinke-Talkner-Dodge-1996-Nature}
\Name{Ghashghaie S., Breymann W., Peinke J., Talkner P. \and Dodge Y.}
  \REVIEW{Nature }{381}{1996}{767}.

\bibitem{Mantegna-Stanley-1996-Nature}
\Name{Mantegna R.~N. \and Stanley H.~E.} \REVIEW{Nature }{383}{1996}{587}.

\bibitem{Mandelbrot-1999-SA}
\Name{Mandelbrot B.~B.} \REVIEW{Sci. Am. }{298}{1999}{70}.

\bibitem{Eisler-Kertesz-Yook-Barabasi-2005-EPL}
\Name{Eisler Z., Kert{\'e}sz J., Yook S.-H. \and Barab{\'a}si A.-L.}
  \REVIEW{Europhys. Lett. }{69}{2005}{664}.

\bibitem{Eisler-Kertesz-2007-EPL}
\Name{Eisler Z. \and Kert{\'e}sz J.} \REVIEW{EPL }{77}{2007}{28001}.

\bibitem{Jiang-Guo-Zhou-2007-EPJB}
\Name{Jiang Z.-Q., Guo L. \and Zhou W.-X.} \REVIEW{Eur. Phys. J. B
  }{57}{2007}{347}.

\bibitem{Vandewalle-Ausloos-1998-EPJB}
\Name{Vandewalle N. \and Ausloos M.} \REVIEW{Eur. Phys. J. B }{4}{1998}{257}.

\bibitem{Ivanova-Ausloos-1999-EPJB}
\Name{Ivanova K. \and Ausloos M.} \REVIEW{Eur. Phys. J. B }{8}{1999}{665}.

\bibitem{Schmitt-Schertzer-Lovejoy-1999-ASMDA}
\Name{Schmitt F., Schertzer D. \and Lovejoy S.} \REVIEW{Appl. Stoch. Models
  Data Anal. }{15}{1999}{29}.

\bibitem{Schmitt-Schertzer-Lovejoy-2000-IJTAF}
\Name{Schmitt F., Schertzer D. \and Lovejoy S.} \REVIEW{Int. J. Theoret. Appl.
  Financ. }{3}{2000}{361}.

\bibitem{Calvet-Fisher-2002-RES}
\Name{Calvet L. \and Fisher A.} \REVIEW{Rev. Econ. Stat. }{84}{2002}{381}.

\bibitem{Ausloos-Ivanova-2002-CPC}
\Name{Ausloos M. \and Ivanova K.} \REVIEW{Comput. Phys. Commun.
  }{147}{2002}{582}.

\bibitem{Gorski-Drozdz-Speth-2002-PA}
\Name{G{\'o}rski A.~Z., Dro{\.{z}}d{\.{z}} S. \and Speth J.} \REVIEW{Physica A
  }{316}{2002}{496}.

\bibitem{AlvarezRamirez-Cisneros-IbarraValdez-Soriano-2002-PA}
\Name{Alvarez-Ramirez J., Cisneros M., Ibarra-Valdez C. \and Soriano A.}
  \REVIEW{Physica A }{313}{2002}{651}.

\bibitem{Balcilar-2003-EMFT}
\Name{Balcilar M.} \REVIEW{Emerging Markets Financ. Trade }{39}{2003}{5}.

\bibitem{Lee-Lee-2005a-JKPS}
\Name{Lee K.~E. \and Lee J.~W.} \REVIEW{J. Korean Phys. Soc. }{46}{2005}{726}.

\bibitem{Lee-Lee-Rikvold-2006-PA}
\Name{Lee J.~W., Lee K.~E. \and Rikvold P.~A.} \REVIEW{Physica A
  }{364}{2006}{355}.

\bibitem{Jiang-Zhou-2007-PA}
\Name{Jiang Z.-Q. \and Zhou W.-X.} \REVIEW{Physica A }{381}{2007}{343}.

\bibitem{Kantelhardt-Zschiegner-KoscielnyBunde-Havlin-Bunde-Stanley-2002-PA}
\Name{Kantelhardt J.~W., Zschiegner S.~A., Koscielny-Bunde E., Havlin S., Bunde
  A. \and Stanley H.~E.} \REVIEW{Physica A }{316}{2002}{87}.

\bibitem{Matia-Ashkenazy-Stanley-2003-EPL}
\Name{Matia K., Ashkenazy Y. \and Stanley H.~E.} \REVIEW{Europhys. Lett.
  }{61}{2003}{422}.

\bibitem{Kwapien-Oswiecimka-Drozdz-2005-PA}
\Name{Kwapie{\'{n}} J., O{\'{s}}wi{\c{e}}cimka P. \and Dro{\.{z}}d{\.{z}} S.}
  \REVIEW{Physica A }{350}{2005}{466}.

\bibitem{Lee-Lee-2005b-JKPS}
\Name{Lee K.~E. \and Lee J.~W.} \REVIEW{J. Korean Phys. Soc. }{47}{2005}{185}.

\bibitem{Oswiecimka-Kwapien-Drozdz-2005-PA}
\Name{O{\'s}wi{\c{e}}cimka P., Kwapie{\'n} J. \and Dro{\.z}d{\.z} S.}
  \REVIEW{Physica A }{347}{2005}{626}.

\bibitem{Jiang-Ma-Cai-2007-PA}
\Name{Jiang J., Ma K. \and Cai X.} \REVIEW{Physica A }{378}{2007}{399}.

\bibitem{Lee-Lee-2007-PA}
\Name{Lee K.~E. \and Lee J.~W.} \REVIEW{Physica A }{383}{2007}{65}.

\bibitem{Lim-Kim-Lee-Kim-Lee-2007-PA}
\Name{Lim G., Kim S., Lee H., Kim K. \and Lee D.-I.} \REVIEW{Physica A
  }{386}{2007}{259}.

\bibitem{Su-Wang-Huang-2009-JKPS}
\Name{Su Z.-Y., Wang Y.-T. \and Huang H.-Y.} \REVIEW{J. Korean Phys. Soc.
  }{54}{2009}{1395}.

\bibitem{Sun-Chen-Wu-Yuan-2001-PA}
\Name{Sun X., Chen H.-P., Wu Z.-Q. \and Yuan Y.-Z.} \REVIEW{Physica A
  }{291}{2001}{553}.

\bibitem{Sun-Chen-Yuan-Wu-2001-PA}
\Name{Sun X., Chen H.-P., Yuan Y.-Z. \and Wu Z.-Q.} \REVIEW{Physica A
  }{301}{2001}{473}.

\bibitem{Ho-Lee-Wang-Chuang-2004-PA}
\Name{Ho D.-S., Lee C.-K., Wang C.-C. \and Chuang M.} \REVIEW{Physica A
  }{332}{2004}{448}.

\bibitem{Wei-Huang-2005-PA}
\Name{Wei Y. \and Huang D.-S.} \REVIEW{Physica A }{355}{2005}{497}.

\bibitem{Gu-Chen-Zhou-2007-EPJB}
\Name{Gu G.-F., Chen W. \and Zhou W.-X.} \REVIEW{Eur. Phys. J. B
  }{57}{2007}{81}.

\bibitem{Du-Ning-2008-PA}
\Name{Du G.-X. \and Ning X.-X.} \REVIEW{Physica A }{387}{2008}{261}.

\bibitem{Zhuang-Yuan-2008-PA}
\Name{Zhuang X.-T. \and Yuan Y.} \REVIEW{Physica A }{387}{2008}{511}.

\bibitem{Wei-Wang-2008-PA}
\Name{Wei Y. \and Wang P.} \REVIEW{Physica A }{387}{2008}{1585}.

\bibitem{Zhou-2010-cnJMSC}
\Name{Zhou W.-X.} \REVIEW{J. Manag. Sci. China (in Chinese) }{13}{2010}{in
  press}.

\bibitem{Jiang-Zhou-2008a-PA}
\Name{Jiang Z.-Q. \and Zhou W.-X.} \REVIEW{Physica A }{387}{2008}{3605}.

\bibitem{Su-Wang-2009-JKPS}
\Name{Su Z.-Y. \and Wang Y.-T.} \REVIEW{J. Korean Phys. Soc. }{54}{2009}{1385}.

\bibitem{Struzik-Siebes-2002-PA}
\Name{Struzik Z.~R. \and Siebes A. P. J.~M.} \REVIEW{Physica A
  }{309}{2002}{388}.

\bibitem{Turiel-Perez-Vicente-2003-PA}
\Name{Turiel A. \and P{\'e}rez-Vicente C.~J.} \REVIEW{Physica A
  }{322}{2003}{629}.

\bibitem{Turiel-Perez-Vicente-2005-PA}
\Name{Turiel A. \and P{\'e}rez-Vicente C.~J.} \REVIEW{Physica A
  }{355}{2005}{475}.

\bibitem{Oswiecimka-Kwapien-Drozdz-Rak-2005-APPB}
\Name{O{\'s}wi{\c{e}}cimka P., Kwapie{\'n} J., Dro{\.z}d{\.z} S. \and Rak R.}
  \REVIEW{Acta Phys. Pol. B }{36}{2005}{2447}.

\bibitem{Zunino-Tabak-Figliola-Perez-Garavaglia-Rosso-2008-PA}
\Name{Zunino L., Tabak B.~M., Figliola A., P{\'e}rez D.~G., Garavaglia M. \and
  Rosso O.~A.} \REVIEW{Physica A }{387}{2008}{6558}.

\bibitem{Bouchaud-Potters-Meyer-2000-EPJB}
\Name{Bouchaud J.-P., Potters M. \and Meyer M.} \REVIEW{Eur. Phys. J. B
  }{13}{2000}{595}.

\bibitem{Saichev-Sornette-2006-PRE}
\Name{Saichev A. \and Sornette D.} \REVIEW{Phys. Rev. E }{74}{2006}{011111}.

\bibitem{Lux-2004-IJMPC}
\Name{Lux T.} \REVIEW{Int. J. Modern Phys. C }{15}{2004}{481}.

\bibitem{Kumar-Deo-2009-PA}
\Name{Kumar S. \and Deo N.} \REVIEW{Physica A }{388}{2009}{1593}.

\bibitem{deSouza-Queiros-2009-CSF}
\Name{de~Souza J. \and Queir{\'o}s S. M.~D.} \REVIEW{Chaos, Solitons \&
  Fractals }{42}{2009}{2512}.

\bibitem{Drozdz-Kwapien-Oswiecimka-Speth-2009-XXX}
\Name{Dro{\.{z}}d{\.{z}} S., Kwapie{\'{n}} J., O{\'{s}}wiecimka P. \and Speth
  J.} \Book{{Quantitative features of multifractal subtleties in time series}}
  arXiv: 0907.2866 (2009).

\bibitem{Nakao-2000-PLA}
\Name{Nakao H.} \REVIEW{Phys. Lett. A }{266}{2000}{282}.

\bibitem{vonHardenberg-Thieberger-Provenzale-2000-PLA}
\Name{von Hardenberg J., Thieberger R. \and Provenzale A.} \REVIEW{Phys. Lett.
  A }{269}{2000}{303}.

\bibitem{Oh-Eom-Havlin-Jung-Wang-Stanley-Kim-2010-PRE}
\Name{Oh G., Eom C., Havlin S., Jung W.-S., Wang F.-Z., Stanley H.~E. \and Kim
  S.} \REVIEW{Phys. Rev. E X}{XX}{2010}{XXX}.

\bibitem{Norouzzadeh-Rahmani-2006-PA}
\Name{Norouzzadeh P. \and Rahmani B.} \REVIEW{Physica A }{367}{2006}{328}.

\bibitem{Hu-Ivanov-Chen-Carpena-Stanley-2001-PRE}
\Name{Hu K., Ivanov P.~C., Chen Z., Carpena P. \and Stanley H.~E.}
  \REVIEW{Phys. Rev. E }{64}{2001}{011114}.

\bibitem{Chen-Ivanov-Hu-Stanley-2002-PRE}
\Name{Chen Z., Ivanov P.~C., Hu K. \and Stanley H.~E.} \REVIEW{Phys. Rev. E
  }{65}{2002}{041107}.

\bibitem{Chen-Hu-Carpena-Bernaola-Galvan-Stanley-Ivanov-2005-PRE}
\Name{Chen Z., Hu K., Carpena P., Bernaola-Galvan P., Stanley H.~E. \and Ivanov
  P.~C.} \REVIEW{Phys. Rev. E }{71}{2005}{011104}.

\bibitem{Halsey-Jensen-Kadanoff-Procaccia-Shraiman-1986-PRA}
\Name{Halsey T.~C., Jensen M.~H., Kadanoff L.~P., Procaccia I. \and Shraiman
  B.~I.} \REVIEW{Phys. Rev. A }{33}{1986}{1141}.

\bibitem{Mandelbrot-1990a-PA}
\Name{Mandelbrot B.~B.} \REVIEW{Physica A }{163}{1990}{306}.

\bibitem{Mandelbrot-1991-PRSA}
\Name{Mandelbrot B.~B.} \REVIEW{Proc. R. Soc. Lond. A }{434}{1991}{79}.

\bibitem{Theiler-Eubank-Longtin-Galdrikian-Farmer-1992-PD}
\Name{Theiler J., Eubank S., Longtin A., Galdrikian B. \and Farmer J.~D.}
  \REVIEW{Physica D }{58}{1992}{77}.

\bibitem{Bogachev-Eichner-Bunde-2007-PRL}
\Name{Bogachev M.~I., Eichner J.~F. \and Bunde A.} \REVIEW{Phys. Rev. Lett.
  }{99}{2007}{240601}.

\bibitem{Zhou-2008-PRE}
\Name{Zhou W.-X.} \REVIEW{Phys. Rev. E }{77}{2008}{066211}.

\bibitem{Gu-Chen-Zhou-2008a-PA}
\Name{Gu G.-F., Chen W. \and Zhou W.-X.} \REVIEW{Physica A }{387}{2008}{495}.

\bibitem{Gopikrishnan-Meyer-Amaral-Stanley-1998-EPJB}
\Name{Gopikrishnan P., Meyer M., Amaral L. A.~N. \and Stanley H.~E.}
  \REVIEW{Eur. Phys. J. B }{3}{1998}{139}.

\bibitem{Schreiber-Schmitz-1996-PRL}
\Name{Schreiber T. \and Schmitz A.} \REVIEW{Phys. Rev. Lett. }{77}{1996}{635}.

\bibitem{Ashkenazy-Ivanov-Havlin-Peng-Goldberger-Stanley-2001-PRL}
\Name{Ashkenazy Y., Ivanov P.~C., Havlin S., Peng C.-K., Goldberger A.~L. \and
  Stanley H.~E.} \REVIEW{Phys. Rev. Lett. }{86}{2001}{1900}.

\bibitem{Ashkenazy-Havlin-Ivanov-Peng-SchulteFrohlinde-Stanley-2003-PA}
\Name{Ashkenazy Y., Havlin S., Ivanov P.~C., Peng C.-K., Schulte-Frohlinde V.
  \and Stanley H.~E.} \REVIEW{Physica A }{323}{2003}{19}.

\bibitem{Ivanov-Amaral-Goldberger-Havlin-Rosenblum-Struzik-Stanley-1999-Nature}
\Name{Ivanov P.~C., Amaral L. A.~N., Goldberger A.~L., Havlin S., Rosenblum
  M.~G., Struzik Z.~R. \and Stanley H.~E.} \REVIEW{Nature }{399}{1999}{461}.

\bibitem{Ivanov-Amaral-Goldberger-Havlin-Rosenblum-Stanley-Struzik-2001-Chaos}
\Name{Ivanov P.~C., Amaral L. A.~N., Goldberger A.~L., Havlin S., Rosenblum
  M.~G., Stanley H.~E. \and Struzik Z.~R.} \REVIEW{Chaos }{11}{2001}{641}.

\bibitem{Ivanov-Rosenblum-Peng-Mietus-Havlin-Stanley-Goldberger-1998-PA}
\Name{Ivanov P.~C., Rosenblum M.~G., Peng C.-K., Mietus J.~E., Havlin S.,
  Stanley H.~E. \and Goldberger A.~L.} \REVIEW{Physica A }{249}{1998}{587}.

\end{thebibliography}

\end{document}